\RequirePackage{fix-cm}
\documentclass[natbib,smallextended]{svjour3}       

\smartqed  

\usepackage{amsmath,amssymb,amsfonts}
\usepackage{algorithmic}
\usepackage{graphicx}
\usepackage{textcomp}

 \usepackage{booktabs}
\usepackage{txfonts}
\usepackage{mathdots}

\usepackage{algorithm}
\usepackage{algorithmic}

  \usepackage{rotating}
        \usepackage{multirow}
        \usepackage{lscape}

\usepackage{color,soul}

\usepackage{xcolor}
\usepackage{caption}

\usepackage{subcaption}

\usepackage{hhline}

 \journalname{}
\begin{document}

\title{ Deep Sentiment Classification and Topic Discovery on Novel Coronavirus or COVID-19 Online Discussions: NLP Using LSTM Recurrent Neural Network Approach
}


\author{Hamed Jelodar $^1$ \and Yongli Wang $^1$ \and  Rita Orji$^2$ \and Hucheng Huang$^3$
}


\institute{Hamed Jelodar \at
              jelodar@njust.edu.cn\\\\
              Yongli Wang\\
              yongliwang@njust.edu.cn\\\\
              Rita Orji\\
              rita.orji@dal.ca\\\\
              Hucheng Huang\\
              schuang6@126.com\\
           \and
             \\
              $^1$ School of Computer Science and Technology, Nanjing University of Science and Technology, Nanjing 210094, China\\
              $^2$ Faculty of Computer Science, Dalhousie University, Halifax, NS, Canada\\
              $^3$ School of Computer, Jiangsu University of Science and Technology, Zhenjiang 212003, China 
                \at      
}

\date{Received: date / Accepted: date}

\maketitle

\begin{abstract}
Internet forums and public social media, such as online healthcare forums, provide a convenient channel for users (people/patients) concerned about health issues to discuss and share information with each other. In late December 2019, an outbreak of a novel coronavirus (infection from which results in the disease named COVID-19) was reported, and, due to the rapid spread of the virus in other parts of the world, the World Health Organization declared a state of emergency. In this paper, we used automated extraction of COVID-19--related discussions from social media and a natural language process (NLP) method based on topic modeling to uncover various issues related to COVID-19 from public opinions. Moreover, we also investigate how to use LSTM recurrent neural network for sentiment classification of COVID-19 comments. Our findings shed light on the importance of using public opinions and suitable computational techniques to understand issues surrounding COVID-19 and to guide related decision-making.


\end{abstract}

\section{Introduction}
Online forums, such as reddit, enable healthcare service providers to collect people/patient experience data. These forums are valuable sources of people's opinions, which can be examined for knowledge discovery and user behaviour analysis. In a typical sub-reddit forum, a user can use keywords and apply search tools to identify relevant questions/answers or comments sent in by other reddit users. Moreover, a registered user can create a topic or post a new question to start discussions with other community members. In answering the questions,  users reflect and share their views and experiences. In these online forums, people may express their positive and negative comments, or share questions, problems, and needs related to health issues. By analysing these comments, we can identify valuable recommendations for improving health-services and understanding the problems of users.\\	

In late December 2019, the outbreak of a novel coronavirus causing COVID-19 was reported [1]. Due to the rapid spread of the virus, the World Health Organization declared a state of emergency. In this paper, we focused on analysing COVID-19--related comments to detect sentiment and semantic ideas relating to COVID-19 based on the public opinions of people on reddit. Specifically, we used automated extraction of COVID-19--related discussions from social media and a natural language process (NLP) method based on topic modeling to uncover various issues related to COVID-19 from public opinions. The main contributions of this paper are as follows:
\begin{itemize}
	\item We present a systematic framework based on NLP that is capable of extracting meaningful topics from COVID-19--related comments on reddit.
	\item We propose a deep learning model based on Long Short-Term Memory (LSTM) for sentiment classification of COVID-19--related comments, which produces better results compared with several other well-known machine-learning methods.
	\item We detect and uncover meaningful topics that are being discussed on COVID-19--related issues on reddit, as primary research. 
	\item We calculate the polarity of the COVID-19 comments related to sentiment and opinion analysis from 10 sub-reddits.
\end{itemize}
Our findings shed light on the importance of using public opinions and suitable computational techniques to understand issues surrounding COVID-19 and to guide related decision-making. Overall, the paper is structured as follows. First, we provide a brief introduction to online healthcare forums. Discussion of COVID-19--related issues and some similar works is provided in section 2. In section 3, we describe the data pre-processing methods adopted in our research, and the NLP and deep-learning methods applied to the COVID-19 comments database. Next, we present the results and discussion. Finally, we conclude and discuss future works based on NLP approaches for analysing the online community in relation to the topic of COVID-19.

\section{Related Work}
Machine and deep-learning approaches based on sentiment and semantic analysis are popular methods of analysing text-content in online health forums. Many researchers have used these methods on social media such as Twitter, reddit [2] - [7], and health information websites [8], [9]. For example; Halder and colleagues [10] focused on exploring linguistic changes to analyse the emotional status of a user over time. They utilized a recurrent neural network (RNN) to investigate user-content in a huge dataset from the mental-health online forums of healthboards.com. McRoy and colleagues [11] investigated ways to automate identification of the information needs of breast cancer survivors based on user-posts of online health forums. Chakravorti and colleagues [12] extracted topics based on various health issues discussed in online forums by evaluating user posts of several subreddits (e.g., r/Depression, r/Anxiety) from 2012 to 2018. VanDam and colleagues [13] presented a classification approach for identifying clinic-related posts in online health communities. For that dataset, the authors collected 9576 thread-initiating posts from WebMD, which is a health information website.

 \begin{figure}
\centering
  \includegraphics[height=7.1cm,width=11cm]{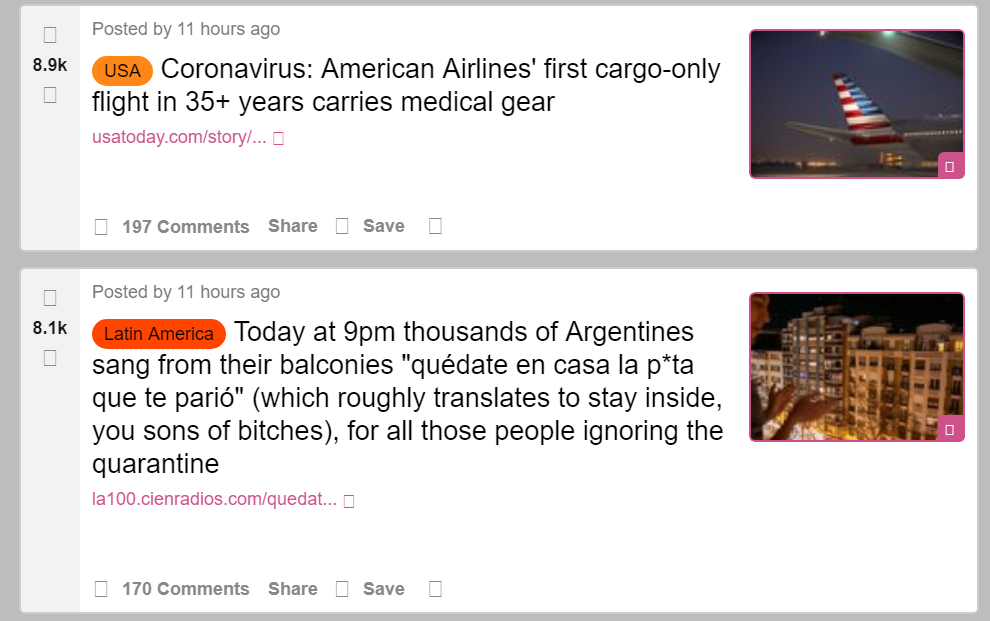}
\caption{Example of user-questions about "COVID-19"on reddit}
\label{fig:1}       
\end{figure}

The COVID-19--related comments from an online healthcare-oriented group can be considered potentially useful for extracting meaningful topics to better understand the opinions and highlight discussions of people/users and improve health strategies. Although there are similar works regarding various health issues in online forums, to the best of our knowledge, this is the first study to utilize NLP methods to evaluate COVID-19--related comments from sub-reddit forums. We propose utilizing the NLP technique based on topic modeling algorithms to automatically extract meaningful topics and design a deep-learning model based on LSTM RNN for sentiment classification on COVID-19 comments and to understand the positive or negative opinions of people as they relate to COVID-19 issues to inform relevant decision-making.

\section{ Framework Methodology}
This section clarifies the methods used to investigate the main contributions to this study, which proposes the use of an unsupervised topic model, with a collaborative deep-learning model based on LSTN RNN to analyse COVID-19--related comments from sub-reddits. The developed framework, shown in Fig. 2, uses sentiment and semantic analysis for mining and opinion analysis of COVID-19--related comments.

\begin{figure}
\centering
  \includegraphics[height=21.14cm,width=15cm]{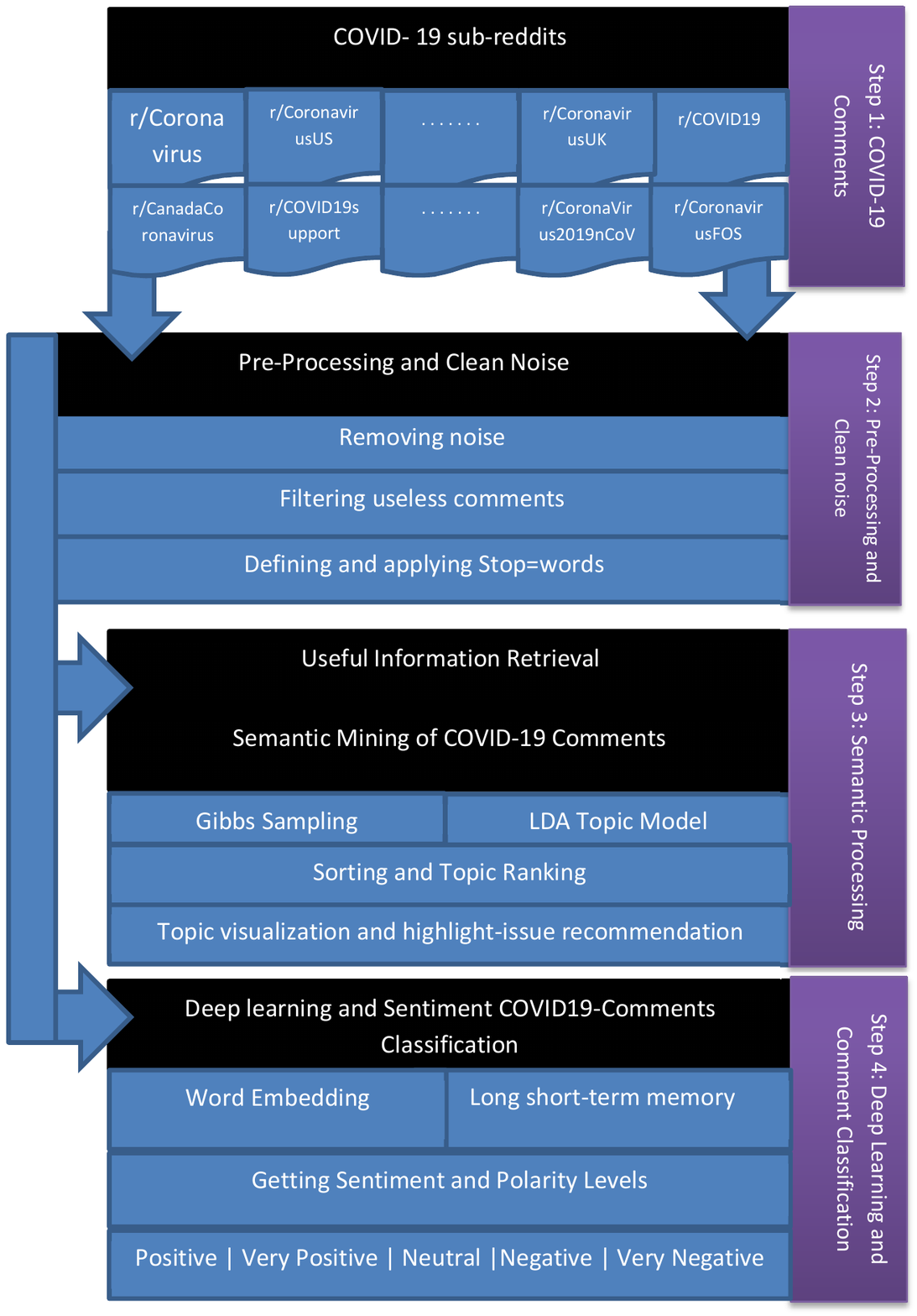}
\caption{ An overview of the research framework for obtaining meaningful results of COVID-19 comments}
\label{fig:1}       
\end{figure}

\subsection{Preparing the input data} 
Reddit is an American social media, a discussion website for various topics that includes web content ratings. In this social media, users are able to post questions and comments, and to respond to each other regarding different subjects, such as COVID-19. The posts are organised by subjects created by online users, called "sub-reddits", which cover a variety of topics like news, science, healthcare, video, books, fitness, food, and image-sharing. This website is an ideal source for collecting health-related information about COVID-19--related issues. This paper focuses on COVID-19--related comments of 10 sub-reddits based on an existing dataset as the first step in producing this model.

\subsection{Removing Noise and Stop-words}
One of the most important steps in pre-processing COVID-19--related comments is removing useless words/data, which are defined as stop-words in NLP, from pure text. Moreover, we also decreased the dimensionality of the features space by eliminating stop-words. For example, the most common words in the text comments are words that are usually meaningless and do not effectively influence the output, such as articles, conjunctions, pronouns, and linking verbs. Some examples include: am, is, are, they, the, these, I, that, and, them.

 \subsection{Semantic Extraction and COVID-19 Comment Mining}
Text-document modeling in NLP is a practical technique that represents an individual document and the set of text-documents based on terms appearing in the text-documents. Topic modeling based on Latent Dirichlet Allocation (LDA) [14] is one type of document modelling approach. As a third step, we utilized topic modeling based on an LDA Topic model and Gibbs sampling [15] for semantic extraction and latent topic discovery of COVID-19--related comments. COVID-19 comments, however, can depend on various subjects that are discussed by reddit users. In this step we can detect and discover these meaningful subjects or topics. Therefore, based on the LDA model, we considered a collection of documents, such as COVID-19--related comments and words, as topics (K), where the discrete topic distributions are drawn from a symmetric Dirichlet distribution. The probability of observed data D was computed and obtained from every COVID-19--related comment in a corpus using the following equation:

\begin{equation} 
p(D\vert \alpha,\beta)=\prod_{d=1}^{M}\int p(\theta_{d}\vert \alpha)\left(\prod_{n=1}^{N_{d}} \sum_{z_{dn}}p(z_{dn}\vert\theta_{d})p(w_{dn}\vert z_{dn},\beta)\right)d\theta_{d} 
\end{equation}

Determined $\alphaup$  parameters of topic Dirichlet prior and also considered parameters of word Dirichlet prior as $\betaup$. \textit{M} is the number of text-documents, and \textit{N} is the vocabulary size. Moreover,(\textit{$\alpha$, $\theta$}) was determined for the corpus-level topic distributions with a pair of Dirichlet multinomials.  (\textit{$\beta$, }$\varphi $) was also determined for the topic-word distributions with a pair of Dirichlet multinomials. In addition, the document-level variables were defined as \textit{$\theta$}${}_{~}$\textit{${}_{d}$}, which may be sampled for each document. The word-level variables  $z_{dn}\ ,w_{d_n}\ $, were sampled in each text-document for each word [14].

\begin{algorithm}
\caption{Pre-processing and removing the noise to prepare the input data}
  \hspace*{\algorithmicindent} \textbf{Input} : A bunch of COVID-19 comments as main document context  \\
    \hspace*{\algorithmicindent} \textbf{Output} : A bunch of text document in string.
 
  \begin{algorithmic}[1]
  
  \STATE d\_i= Get\_data(); getting COVID-19 comments as pure data.

      \FOR{d\_i.row (all\_record) != last\_record}
        \STATE   d\_i\_2= d\_i.cleanData(d\_i); removing stop-words, clean noise
         \STATE   d\_i\_2=d\_i\_2.arranged(); processing to arrange dataset.
           
      \ENDFOR
        \RETURN d\_i\_2 as a string

  \end{algorithmic}
\end{algorithm}

\begin{algorithm}
\caption{General Process for Semantic-Comment-Mining via Topic Model}
  \hspace*{\algorithmicindent} \textbf{Input} : A group of COVID-19--related comments as main document context \\
    \hspace*{\algorithmicindent} \textbf{Output} : A set of topics from the documents as integer values 
 
  \begin{algorithmic}[1]
  
  \STATE Pre-process and removing noise and clean data by Algorithm 1.

      \FOR{each topic $k \in \{1,2,\dots,k\}$}
        \STATE  word-probability under the topic of sampling || or the word distribution for topic $k$ among COVID-19--related comments
         \STATE   $\phi  \sim Dirichlet(\beta)$
           
      \ENDFOR

      \FOR{each COVID-19--related comments $d \in \{1,\dots,D\}$}
        \STATE  The topic distribution for document $m$ 
         \STATE $d \theta \sim Dirichlet(\alpha)$
           
            \FOR{ Per word in COVID-19--related content-document $d$}
              \STATE  sampling the distribution of topics in the COVID-19--related comments-documents to obtain the topic of the word:\\ .$Z_{d}\sim Mul(\theta)$
                 \STATE word-sampling undert the topic, $ W_{d} \sim Mul(\phi) $

             \ENDFOR
 
      \ENDFOR

  \end{algorithmic}
\end{algorithm}

Algorithm 2 describes a general process as part of our framework for extracting latent topics and semantic mining. The input data consists of the number of COVID-19--related comments as the context of the document: Line 1 processes the pure-data to eliminate noise and stop-words based on Algorithm 1. Lines 2-5 compute the probability of the word distribution from Topic K[i]. Lines 6-11 compute the probability of the topic distribution from the COVID-19-Content-Document m [i]. As highlighted in Equation 1, the variables $\theta_{m},  w_{n}$ are computed for document-level and word-level of the framework. In more detail, the LDA handles topics as multinomial distributions in documents and words as a probabilistic mixture of a pre-determined number from latent topics. Lines 1-3 of Algorithm 3 show the semantic mining to extract the latent topics. We then used a sorting function to determine the recommended highlighted topics. Because the Gibbs sampling method is used in this step, the time requested for model inference can be specified as the sum of the time for inferring LDA. Therefore, the time complexity for LDA is O(N K), where N denotes the total size of the corpus (COVID-19--related comments) and K is the topic number.

\begin{algorithm}
\caption{COVID-19--Related Comments Mining and Topic Recommendation}
  \hspace*{\algorithmicindent} \textbf{Input} : Importing latent-topics through Algoritm 2   \\
    \hspace*{\algorithmicindent} \textbf{Output} : Recommended top highlight topics of various aspects of COVID-19 comments
 
  \begin{algorithmic}[1]
  
  \STATE Extract semantic contents, trining the LDA Topic Model 
  \STATE Detmining the top topics recommended based on the value of the topic probality of all data.
  \STATE Ranking and sorting the most meaningful topics recommended of COVID-19 comments

        \RETURN A list of recommended highlight topics

  \end{algorithmic}
\end{algorithm}

\subsection{Deep-Learning and Sentiment Classification}
Deep neural networks have been successfully employed for different types of machine-learning tasks, such as NLP-based methods utilizing sentiment aspects for deep classification [16] - [21]. Deep neural networks are able to model high-level abstractions and to decrease the dimensions by utilizing multiple processing layers based on complex structures or to be combined with non-linear transformations. RNNs are popular models with demonstrated importance and strength in most NLP works [22] - [24]. The purpose of RNNs is to use consecutive information, and the output is augmented by storing previous calculations. In fact, RNNs are equipped with a memory function that saves formerly calculated information. Basic RNNs, however, have some challenges due to gradient vanishing or exploding, and they are unable to learn long-term dependencies. LSTM [25], [26] units have the benefit of being able to avoid this challenge by adjusting the information in a cell state using 3 different gates. The formula for each LSTM cell can be formalized as:

\begin{equation}
f_{t}=\sigma(W_{fz}z_{t-1}+ W_{fx}x_{t} + b_{f}  )
\end{equation}

\begin{equation}
i_{t}=\sigma(W_{iz}z_{t-1}+ W_{ix}x_{t} + b_{i}  )
\end{equation}

\begin{equation}
o_{t}=\sigma(W_{oz}z_{t-1}+ W_{ox} x_{t}+b_{o})
\end{equation}

The  forget $(f_{t})$, input $(i_{t})$, and output $(o_{t})$ gates for each LSTM cell are determined by these 3 equations, eqs. 2-4, respectively. In an LSTM layer, the forget gate determines which previous information from the cell state is forgotten. The input gate controls or determines the new information that is saved in the memory cell. The output gate controls or determines the amount of information in the internal memory cell to be exposed. The cell-memory/input block equations are:

\begin{equation}
 \widetilde{C}_{t}=\phi(W_{cz}z_{t-1} + W_{cx} x_{t} + b_{c})
\end{equation}

\begin{equation}
C_{t}=i_{t} \odot  \widetilde{C}_{t} + f_{t} \odot  {C}_{t-1}
\end{equation}

\begin{equation}
 z_{t}=o_{t}\odot\phi(C_{t})
\end{equation}

In which, $C_{i}$ is the cell state, $z_{t}$ is the hidden output, and  $x_{t}$  is an input vector.   $W$ and $b$ are the weight matrix and the bias term respectively. $ \sigma $  is sigmoid and  $ \phi $   is tanh.  $ \odot $  is element-wise multiplication.

As the last step of this framework, an LSTM model was utilised to assess the COVID-19--related comments of online users who posted on reddit, in order to recognize the emotion/sentiment elicited from these comments. We designed two LSTM-layers and for pre-trained embeddings, considered the Glove-50 dimension \footnote{https://www.kaggle.com/watts2/glove6b50dtxt}, which were trained over a large corpus of COVID-19--related comments (Figure 3). The processed text from the COVID-19--related comments, however, is changed to vectors with a fixed dimension by converting pre-trained embeddings. Moreover, COVID-19 comments can also be described as a characters-sequence with its corresponding dimension creating a matrix [27].

\begin{figure}
\centering
  \includegraphics[height=9.14cm,width=11cm]{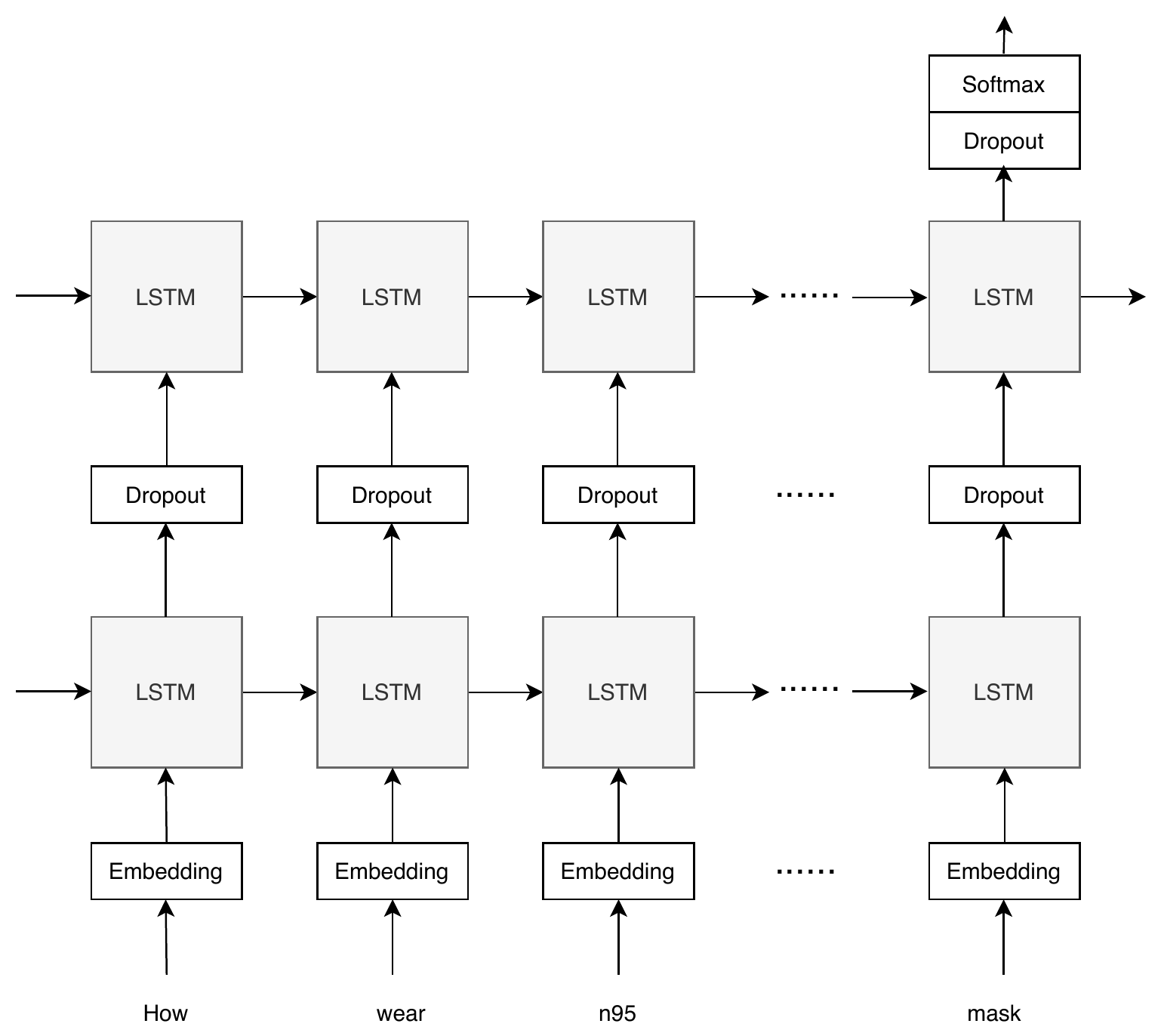}
\caption{Structure of the LSTM designed for COVID-19 sentiment classification.}
\label{fig:1}       
\end{figure}

\section{ Experiment Details}

In this section, we provide a detailed description of the data collection and experimental results followed by a comprehensive discussion of the results. We assessed 563,079 COVID-19--related comments from reddit. The dataset was collected between January 20, 2020 and March 19, 2020 (the full dataset is available at Kaggle website\footnote{ https://www.kaggle.com/khalidalharthi/coronavirus-posts-in-reddit-platform}). We used MALLET\footnote{http://mallet.cs.umass.edu/} to implement the inference and capture the LDA topic model to retrieve latent topics. We used the Python library Keras \footnote{https://pypi.org/project/Keras/} to implement our deep-learning model.

\begin{landscape}
\begin{table}
\centering
\caption{Top 10 topics from COVID-19--related comments on reddit.}
\resizebox{19cm}{!} {
\begin{tabular}{|l|l|c|c|c|c|c|l|l|l|} 
\hline
\multicolumn{10}{|l|}{}                                                                                                                                                                                                                                                                                                                                                                                                                                                                                                                                                                                                                                                                                                                                                                                                                                                                                                                                                                                                                                                                                                                                                                                                                                                                                                                                                                                                                         \\ 
\hline
\multicolumn{2}{|l|}{Topic 85~}                                                                                                                                                                                                                                   & \multicolumn{2}{c|}{Topic 69~}                                                                                                                                                                                                                                                            & \multicolumn{2}{c|}{Topic 8~}                                                                                                                                                                                                                                                               & \multicolumn{2}{c|}{Topic 18~}                                                                                                                                                                                                                                                                   & \multicolumn{2}{l|}{Topic 48~}                                                                                                                                                                                                                                                 \\ 
\hline
\multicolumn{2}{|l|}{~Rank 1}                                                                                                                                                                                                                                     & \multicolumn{2}{l|}{Rank 2}                                                                                                                                                                                                                                                               & \multicolumn{2}{l|}{Rank 3}                                                                                                                                                                                                                                                                 & \multicolumn{2}{l|}{Rank 4}                                                                                                                                                                                                                                                                      & \multicolumn{2}{l|}{~Rank 5}                                                                                                                                                                                                                                                   \\ 
\hline
\multicolumn{2}{|l|}{Propration : 12.7966}                                                                                                                                                                                                                        & \multicolumn{2}{l|}{Propration~:~ 6.90415}                                                                                                                                                                                                                                                & \multicolumn{2}{l|}{Propration~:~ ~5.72494}                                                                                                                                                                                                                                                 & \multicolumn{2}{l|}{Propration~:~5.5769}                                                                                                                                                                                                                                                         & \multicolumn{2}{l|}{Propration~:~ 5.28395}                                                                                                                                                                                                                                     \\ 
\hline
\begin{tabular}[c]{@{}l@{}} people\\virus \\day \\bad\\~stop \\news \\worse \\days \\big \\understand\end{tabular}               & \begin{tabular}[c]{@{}l@{}}sick \\great \\spread \\start \\person \\told \\contact \\family \\spreading \\coming~\end{tabular} & \begin{tabular}[c]{@{}c@{}} china \\population\\~means \\hard \\years \\question \\place \\comment \\kind \\average \end{tabular}       & \begin{tabular}[c]{@{}c@{}}normal~\\found~\\general~\\~clear \\takes \\real \\start \\single \\similar \\simply\end{tabular}                    & \begin{tabular}[c]{@{}c@{}} people \\die \\shit \\fuck\\~long \\life \\care \\wrong \\money \\fucking \end{tabular}                     & \begin{tabular}[c]{@{}c@{}}free \\times \\happen \\lives \\hate \\save \\governments \\economy \\dying \\imagine~\end{tabular}                    & \begin{tabular}[c]{@{}c@{}} virus \\people \\symptoms\\~infection \\cases \\disease \\pneumonia \\case \\coronavirus\\infected~\end{tabular}         & \begin{tabular}[c]{@{}l@{}}severe \\risk \\source \\long \\pretty \\infections \\treatment \\viruses \\information \\article\end{tabular} & \begin{tabular}[c]{@{}l@{}} people \\testing\\~government \\country \\tested \\test \\infected \\home \\covid \\pandemic \end{tabular}     & \begin{tabular}[c]{@{}l@{}}countries \\care \\symptoms \\tests \\spread \\situation \\south \\social\\shut numbers~\end{tabular}  \\ 
\hline
\multicolumn{2}{|l|}{Topic 9~}                                                                                                                                                                                                                                    & \multicolumn{2}{c|}{Topic 30~}                                                                                                                                                                                                                                                            & \multicolumn{2}{c|}{~Topic 58~}                                                                                                                                                                                                                                                             & \multicolumn{2}{l|}{~Topic 76~}                                                                                                                                                                                                                                                                  & \multicolumn{2}{l|}{Topic 63~}                                                                                                                                                                                                                                                 \\ 
\hline
\multicolumn{2}{|l|}{Rank 6~}                                                                                                                                                                                                                                     & \multicolumn{2}{c|}{Rank 7~}                                                                                                                                                                                                                                                              & \multicolumn{2}{l|}{~Rank 8}                                                                                                                                                                                                                                                                & \multicolumn{2}{c|}{Rank 9}                                                                                                                                                                                                                                                                      & \multicolumn{2}{c|}{Rank 10}                                                                                                                                                                                                                                                   \\ 
\hline
\multicolumn{2}{|l|}{~Propration~:~ ~5.03657}                                                                                                                                                                                                                     & \multicolumn{2}{l|}{Propration~:~ 4.75303}                                                                                                                                                                                                                                                & \multicolumn{2}{l|}{\begin{tabular}[c]{@{}l@{}}Propration~:~~\\ 4.62488\end{tabular}}                                                                                                                                                                                                       & \multicolumn{2}{l|}{Propration~:~~4.41009}                                                                                                                                                                                                                                                       & \multicolumn{2}{l|}{Propration~:~ ~0.36916}                                                                                                                                                                                                                                    \\ 
\hline
\begin{tabular}[c]{@{}l@{}} good \\thinking \\working\\~stuff \\bit \\happen \\small \\works \\experience \\future \end{tabular} & \begin{tabular}[c]{@{}l@{}}group \\home \\worried \\month \\expect \\support \\side \\heard chance bring~\end{tabular}         & \multicolumn{1}{l|}{\begin{tabular}[c]{@{}l@{}} good \\hope \\feel \\house\\~started \\safe \\fine \\hard \\months \\live\end{tabular}} & \multicolumn{1}{l|}{\begin{tabular}[c]{@{}l@{}}~friend \\wife \\healthy \\times \\kind \\hit \\doctor \\person \\coming starting~\end{tabular}} & \multicolumn{1}{l|}{\begin{tabular}[c]{@{}l@{}} home\\stay \\health\\~italy \\today \\cases \\weeks \\risk \\days \\hope \end{tabular}} & \multicolumn{1}{l|}{\begin{tabular}[c]{@{}l@{}}taking \\open \\public \\day \\face \\yesterday \\food \\confirmed \\social \\pretty\end{tabular}} & \multicolumn{1}{l|}{\begin{tabular}[c]{@{}l@{}} health \\idea \\medical\\~months \\wrong \\true \\positive \\travel \\edit \\disease ~\end{tabular}} & \begin{tabular}[c]{@{}l@{}}matter \\correct \\thread \\science \\kids \\result \\majority \\effective \\scale \\specifically\end{tabular} & \begin{tabular}[c]{@{}l@{}} hospital \\medical\\~hospitals \\healthcare \\patients \\care \\public \\city \\health \\person ~\end{tabular} & \begin{tabular}[c]{@{}l@{}}patient \\workers \\staff \\case \\cities \\sick \\room \\beds \\states \\emergency\end{tabular}       \\
\hline
\end{tabular}
}
\end{table}
\end{landscape}

\begin{table}
\centering
\caption{Topics ranking 11 to 25 from COVID-19--related comments on reddit.}
\resizebox{11cm}{!} {
\begin{tabular}{|l|l|l|l|l|l|l|l|l|l|} 
\hline
\multicolumn{10}{|l|}{}                                                                                                                                                                                                                                                                                                                                                                                                                                                                                                                                                                                                                                                                                                                                                                                                                                                                                                                                                                                                                                                                                                                                                                                                                                                                                                                                                                                                                                                                                                                                                                                                                                                                                                          \\ 
\hline
\multicolumn{2}{|l|}{Topic 31}                                                                                                                                                                                                                                                                                                     & \multicolumn{2}{c|}{Topic 17}                                                                                                                                                                                                                                                                                                                 & \multicolumn{2}{c|}{Topic 61 }                                                                                                                                                                                                                                                                                                           & \multicolumn{2}{c|}{Topic 1}                                                                                                                                                                                                                                                                                                                                     & \multicolumn{2}{l|}{Topic 96}                                                                                                                                                                                                                                                                                 \\ 
\hline
\multicolumn{2}{|l|}{ Rank 11}                                                                                                                                                                                                                                                                                     & \multicolumn{2}{l|}{Rank 12}                                                                                                                                                                                                                                                                                                                  & \multicolumn{2}{l|}{Rank 13}                                                                                                                                                                                                                                                                                                                             & \multicolumn{2}{l|}{Rank 14}                                                                                                                                                                                                                                                                                                                                     & \multicolumn{2}{l|}{ Rank 15}                                                                                                                                                                                                                                                                 \\ 
\hline
\multicolumn{2}{|l|}{Propration :  3.81556}                                                                                                                                                                                                                                                                        & \multicolumn{2}{l|}{Propration :  3.53217}                                                                                                                                                                                                                                                    & \multicolumn{2}{l|}{Propration :   3.34602}                                                                                                                                                                                                                                                              & \multicolumn{2}{l|}{Propration : 3.15658}                                                                                                                                                                                                                                                                                        & \multicolumn{2}{l|}{Propration :  2.92918}                                                                                                                                                                                                                    \\ 
\hline
\begin{tabular}[c]{@{}l@{}} cases\\ rate \\flu \\deaths\\ numbers \\death \\days \\china \\case \\infected \end{tabular} & \begin{tabular}[c]{@{}l@{}}mortality \\weeks \\confirmed \\population \\wuhan \\good \\patients \\reported \\fatality \\spread  \end{tabular}           & \multicolumn{1}{c|}{\begin{tabular}[c]{@{}c@{}}coronavirus\\ quarantine \\years \\wait \\stupid \\happening \\shit \\word \\watch \\dangerous \end{tabular}} & \multicolumn{1}{c|}{\begin{tabular}[c]{@{}c@{}}shut \\corona \\national \\words \\mind \\bet \\heard \\source \\sound \\common  \end{tabular}} & \multicolumn{1}{c|}{\begin{tabular}[c]{@{}c@{}} virus \\flu \\country\\ china \\countries \\sars \\pandemic \\chinese \\infected \\bad \end{tabular}} & \multicolumn{1}{c|}{\begin{tabular}[c]{@{}c@{}}test\\confirmed \\control \\spread \\singapore \\human \\epidemic \\global \\remember \\subreddit  \end{tabular}} & \multicolumn{1}{c|}{\begin{tabular}[c]{@{}c@{}} pretty \\shit \\america\\ american \\stay \\love \\reddit \\americans \\usa \\death   \end{tabular}}                   & \begin{tabular}[c]{@{}l@{}}food\\ literally \\month \\guys \\real \\live \\life \\full \\realize \\dude \end{tabular}                   & \begin{tabular}[c]{@{}l@{}} weeks \\measures \\stop\\ panic \\school \\link \\close \\early \\public \\lockdown \end{tabular} & \begin{tabular}[c]{@{}l@{}}yesterday \\action \\economy \\spread \\absolutely \\closed \\country \\expect \\moment \\fine  \end{tabular}      \\ 
\hline
\multicolumn{2}{|l|}{Topic 93}                                                                                                                                                                                                                                                                                                     & \multicolumn{2}{c|}{Topic 4}                                                                                                                                                                                                                                                                                                                  & \multicolumn{2}{c|}{ Topic 40 }                                                                                                                                                                                                                                                                          & \multicolumn{2}{l|}{~Topic 43}                                                                                                                                                                                                                                                                                                                                   & \multicolumn{2}{l|}{Topic 83}                                                                                                                                                                                                                                                                                 \\ 
\hline
\multicolumn{2}{|l|}{Rank 16 }                                                                                                                                                                                                                                                                                     & \multicolumn{2}{c|}{Rank 17 }                                                                                                                                                                                                                                                                                                 & \multicolumn{2}{l|}{ Rank 18}                                                                                                                                                                                                                                                                                                            & \multicolumn{2}{c|}{Rank 19}                                                                                                                                                                                                                                                                                                                                     & \multicolumn{2}{c|}{Rank 20}                                                                                                                                                                                                                                                                                  \\ 
\hline
\multicolumn{2}{|l|}{ Propration :  2.22481}                                                                                                                                                                                                                                       & \multicolumn{2}{l|}{Propration :  2.1929}                                                                                                                                                                                                                                                     & \multicolumn{2}{l|}{Propration :  2.09387}                                                                                                                                                                                                                                                               & \multicolumn{2}{l|}{Propration : 1.91949}                                                                                                                                                                                                                                                                                                                        & \multicolumn{2}{l|}{Propration : 1.73082}                                                                                                                                                                                                                                                                     \\ 
\hline
\begin{tabular}[c]{@{}l@{}} covid \\young \\risk\\ fever \\immune \\age \\sick \\cough \\life \\cold \end{tabular}                       & \begin{tabular}[c]{@{}l@{}} elderly \\older \\years \\healthy \\long \\die \\younger \\coronavirus \\asthma \\conditions  \end{tabular} & \begin{tabular}[c]{@{}l@{}} pay \\money \\companies\\ company \\insurance \\paid \\paying \\free \\cost \\afford \end{tabular}                               & \begin{tabular}[c]{@{}l@{}}tax \\years \\employees \\leave \\bill \\taxes \\healthcare \\job \\business \\income  \end{tabular}                & \begin{tabular}[c]{@{}l@{}} fucking \\fuck\\ government \\guy \\man \\stupid \\fucked \\sick \\rate \\takes \end{tabular}                             & \begin{tabular}[c]{@{}l@{}}lmao \\supposed \\treatment \\ass \\tested \\dumb \\sell \\canada \\damn \\idiots  \end{tabular}                                      & \begin{tabular}[c]{@{}l@{}}   trump \\cdc \\president\\ government \\covid \\country \\coronavirus \\administration \\response \\america \end{tabular} & \begin{tabular}[c]{@{}l@{}}states \\corona \\federal \\pandemic \\united \\pence \\americans \\hoax \\national \\million  \end{tabular} & \begin{tabular}[c]{@{}l@{}}home \\sick \\working\\ workers \\store \\stay \\business \\employees \\job \\weeks \end{tabular}  & \begin{tabular}[c]{@{}l@{}}stores \\essential \\company \\businesses \\told \\grocery \\jobs \\paid \\restaurants \\customers  \end{tabular}  \\ 
\hhline{|==========|}
\multicolumn{2}{|l|}{Topic 86}                                                                                                                                                                                                                                                                                                     & \multicolumn{2}{c|}{Topic 12}                                                                                                                                                                                                                                                                                                                 & \multicolumn{2}{c|}{~Topic 44}                                                                                                                                                                                                                                                                                                                           & \multicolumn{2}{l|}{Topic 99}                                                                                                                                                                                                                                                                                                                                    & \multicolumn{2}{l|}{Topic 88}                                                                                                                                                                                                                                                                                 \\ 
\hline
\multicolumn{2}{|l|}{Rank 21}                                                                                                                                                                                                                                                                                                      & \multicolumn{2}{c|}{Rank 22}                                                                                                                                                                                                                                                                                                                  & \multicolumn{2}{l|}{Rank 23}                                                                                                                                                                                                                                                                                                                             & \multicolumn{2}{c|}{Rank 24}                                                                                                                                                                                                                                                                                                                                     & \multicolumn{2}{c|}{Rank 25}                                                                                                                                                                                                                                                                                  \\ 
\hline
\multicolumn{2}{|l|}{Propration : 1.59421}                                                                                                                                                                                                                                                                                         & \multicolumn{2}{l|}{Propration : 1.54827}                                                                                                                                                                                                                                                                                                      & \multicolumn{2}{l|}{Propration : 1.53468}                                                                                                                                                                                                                                                                                                                & \multicolumn{2}{l|}{Propration : 1.48268}                                                                                                                                                                                                                                                                                                                        & \multicolumn{2}{l|}{Propration : 1.44486}                                                                                                                                                                                                                                                                     \\

\hline
\begin{tabular}[c]{@{}l@{}}~masks \\mask\\~wearing \\wear \\face \\surgical \\protect \\protection \\effective \\infected~\end{tabular}                  & \begin{tabular}[c]{@{}l@{}}medical \\sick \\public \\supply \\shortage \\hands \\prevent \\workers \\gloves \\doctors\end{tabular}                                      & \begin{tabular}[c]{@{}l@{}} test\\~testing \\tested \\tests \\symptoms \\positive \\flu \\cdc \\kits \\fever \end{tabular}                                                   & \begin{tabular}[c]{@{}l@{}}market \\confirmed \\numbers \\free \\negative \\cases \\day \\doctor \\cough \\labs~\end{tabular}                                  & \begin{tabular}[c]{@{}l@{}} market\\stock \\economy \\buy \\years \\supply \\economic \\deaths \\production \\markets \end{tabular}                                   & \begin{tabular}[c]{@{}l@{}}war\\global \\buying \\parts \\chain \\stocks \\manufacturing \\goods \\panic \\run~\end{tabular}                                                     & \begin{tabular}[c]{@{}l@{}} school\\kids \\schools \\flu \\home \\parents \\family \\close \\sick \\children ~\end{tabular}                                                                            & \begin{tabular}[c]{@{}l@{}}closed\\weeks \\care \\husband \\travel \\students \\cdc \\closing \\told \\stay\end{tabular}                                & \begin{tabular}[c]{@{}l@{}} food\\buy\\water \\stock \\buying \\supplies \\rice \\bought \\weeks ~\\eat~\end{tabular}                         & \begin{tabular}[c]{@{}l@{}}store \\days \\stuff \\supply \\panic \\stocked \\family \\canned \\prepping \\prepared\end{tabular}                               \\ 
\hhline{|==========|}

\end{tabular}
}
\end{table}

\begin{landscape}

\begin{figure}
\centering
  \includegraphics[height=10.14cm,width=19cm]{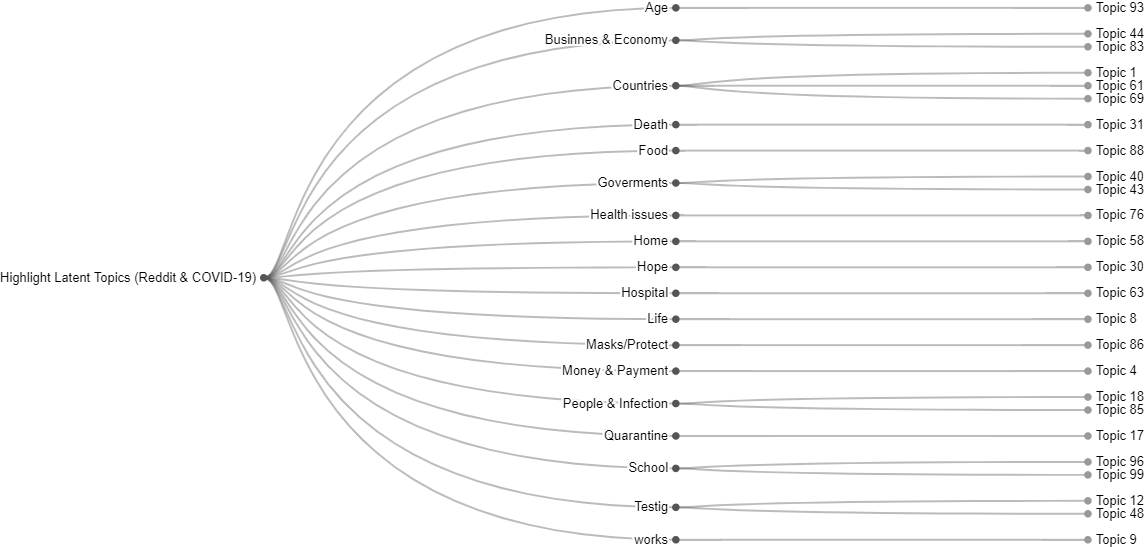}
\caption{Cluster dendrogram of highlight latent topics generated in a COVID-19--related discussion}
\label{fig:1}       
\end{figure}

\end{landscape}

\begin{figure}[h!]
  \centering
  \begin{subfigure}[b]{0.6\linewidth}
    \includegraphics[width=\linewidth]{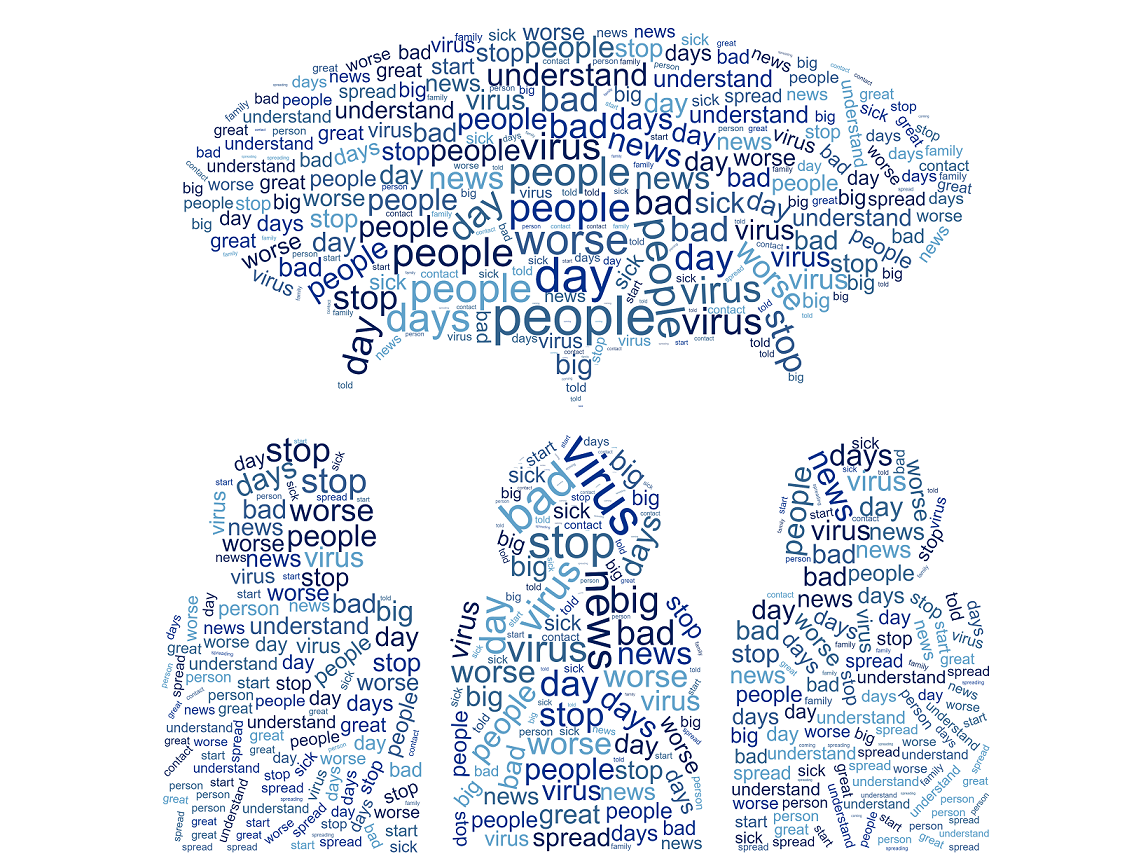}
    \caption{Topic Word 85}
  \end{subfigure}
  \begin{subfigure}[b]{0.6\linewidth}
    \includegraphics[width=\linewidth]{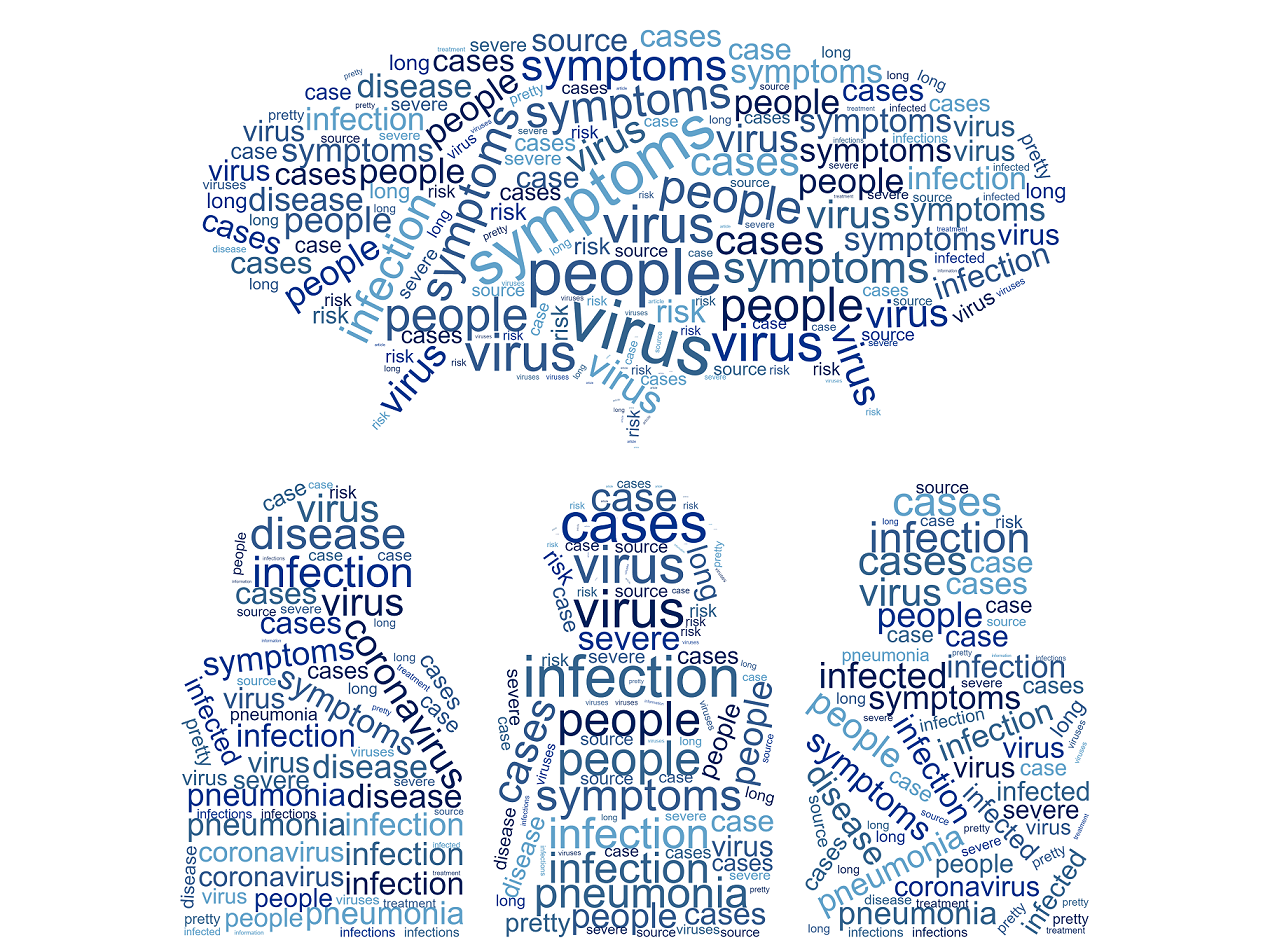}
    \caption{Topic Word 18 }
  \end{subfigure}
  \caption{Word cloud visualisation based on the word-weight of the topics.}
  \label{fig:coffee}
\end{figure}

According to Table 1 and 2 and Figures 4-8, the following observations were made: Topics 85 and 18 had a similar concept in “People/Infection”. Topic 85 included words referring to people, such as "people", "virus", "day", "bad", "stop", "news", "worse", "sick", "spread", and "family". This topic is the first ranked topic discovered from the generated latent topics, in which most users express their opinion and comment on this issue. Based on Table 1 and Figure 5 (a) in this topic, the terms “people” and “virus” were the most highlighted words, with word-weights of 0.1295\% and 0.0301\%, respectively. Also, we can see the importance of the term "family" from this topic. In addition, Topic 18 contains the telling words "virus", "people", "symptoms", "infection", "cases", "disease", "pneumonia", "coronavirus", and "treatment". Other revealing words in Topic 18 included "people", "infection", and "treatment". These terms initially suggest a set of user comments about treatment issues. Moreover, the sentiment analysis of the terms suggest that negative words were more highlighted than positive words.

\begin{figure}[h!]
  \centering
  \begin{subfigure}[b]{0.6\linewidth}
    \includegraphics[width=\linewidth]{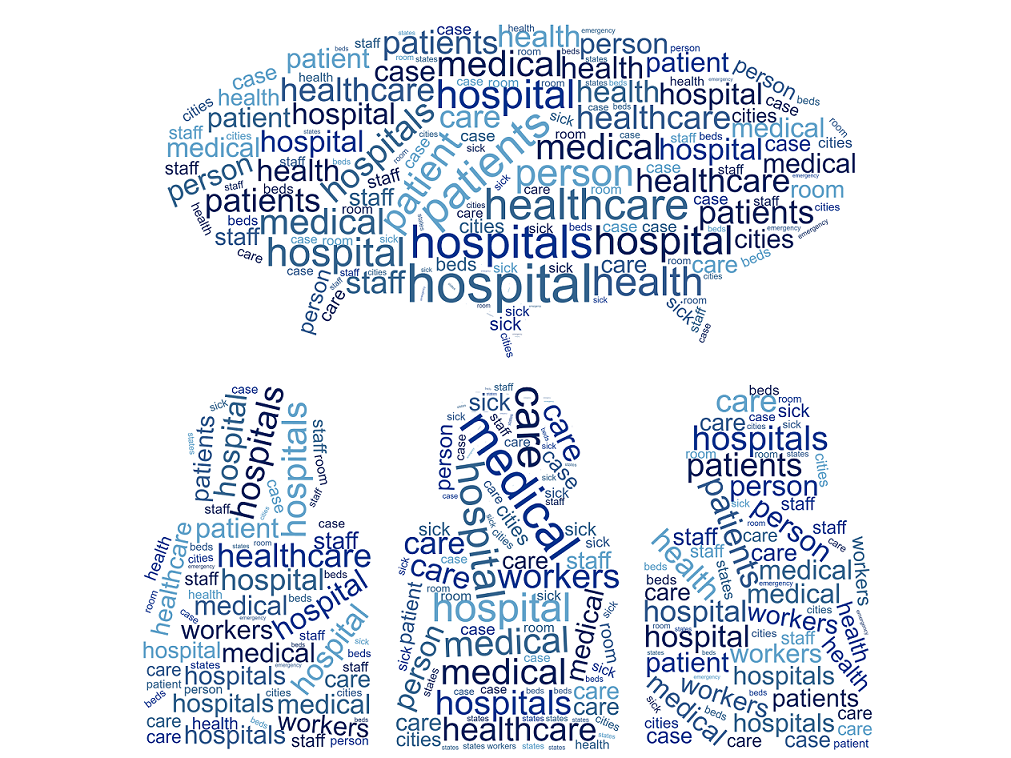}
    \caption{Topic Word 63}
  \end{subfigure}
  \begin{subfigure}[b]{0.6\linewidth}
    \includegraphics[width=\linewidth]{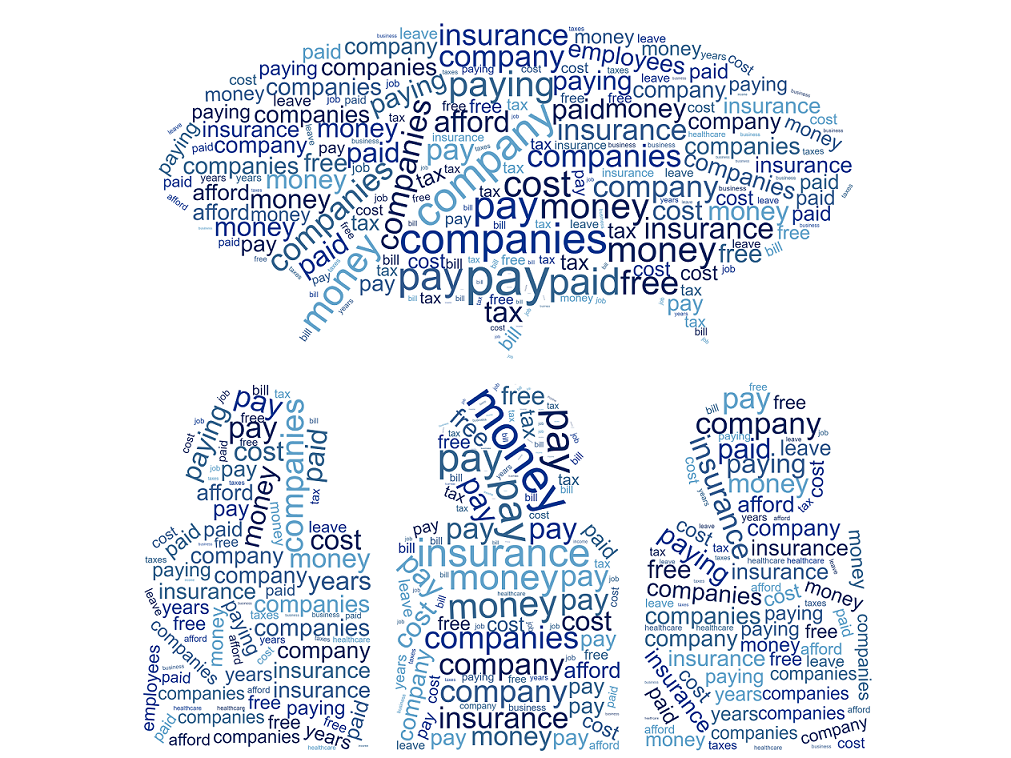}
    \caption{Topic Word 4}
  \end{subfigure}
  \caption{Word cloud visualisation based on the word-weight of the topics.}
  \label{fig:coffee}
\end{figure}

 Topic 63 also addresses healthcare and hospital issues with the most frequent term being “hospital”. Words such as "hospital", "medical", "healthcare", "patients", "care", and "city" were included. The terms “hospital”, “medical”, and “healthcare” were the most highlighted words, with word-weights of 0.0561\%, 0.0282\%, and 0.0278\%, respectively. Other words worth mentioning that were seen for this topic were “person”, “patient", “staff”, “workers”, and “emergency”. Topic 63 was assigned as medical staff issues. Topic 4 included words relating to money, such as "pay", "money", "companies", "insurance", "paid", "free", "cost", "tax", "years", and "employees". Moreover, the sentiment analysis of the terms suggested that negative words were more highlighted than positive words.\\

\begin{figure}[h!]
  \centering
  \begin{subfigure}[b]{0.6\linewidth}
    \includegraphics[width=\linewidth]{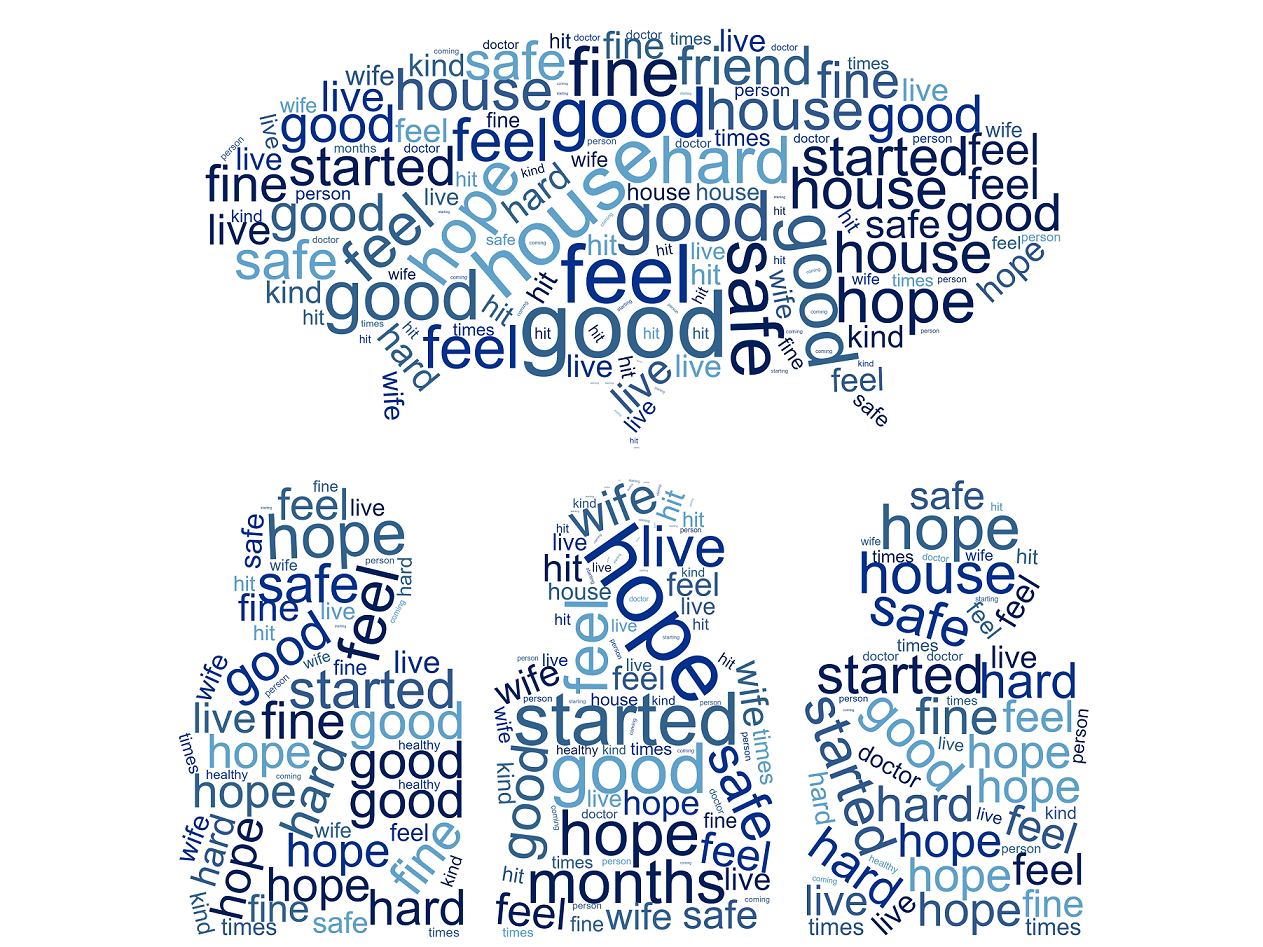}
    \caption{Topic Word 30}
  \end{subfigure}
  \begin{subfigure}[b]{0.6\linewidth}
    \includegraphics[width=\linewidth]{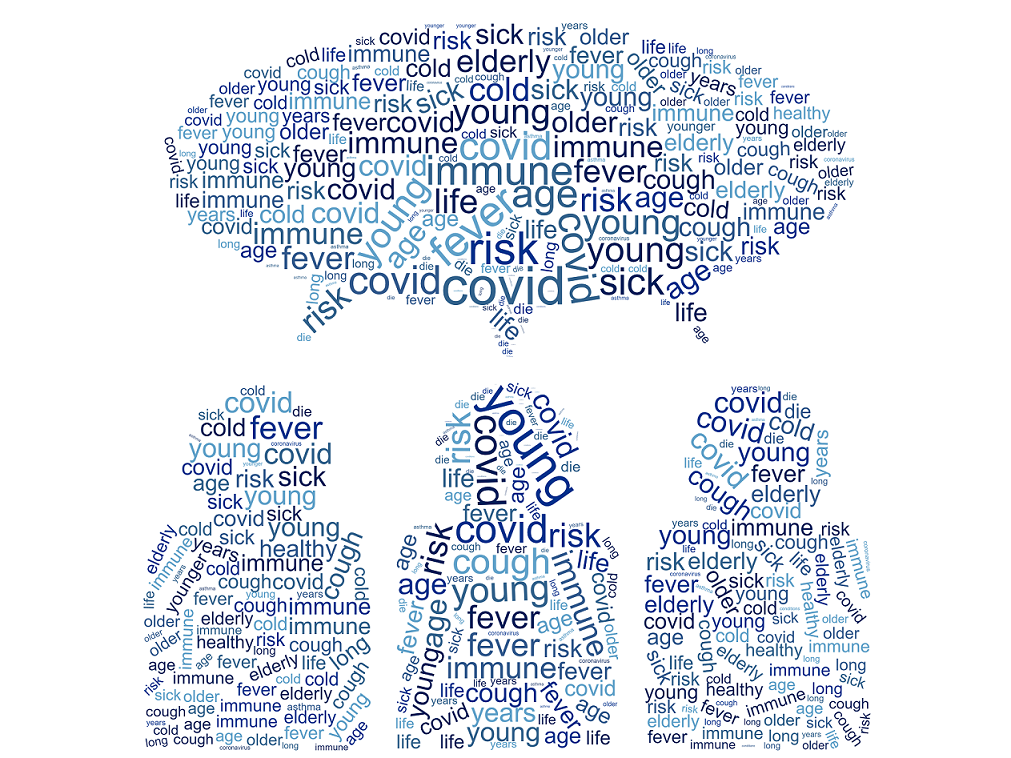}
    \caption{Topic Word 93}
  \end{subfigure}
  \caption{Word cloud visualisation based on the word-weight of the topics.}
  \label{fig:coffee}
\end{figure}

Topic 30 covers user’s comments concerning issues related to "feelings and hopes" and highlight words such as "good", "hope", "feel", "house", "safe", "hard", "months", "fine", "live", and "friend". Moreover, sentiment analysis of terms suggested that positive words were more highlighted than negative words. Positive words such as “good”, “hope”, “safe”, “fine”, “kind”, and “friend”, thus pertain to the phenomenon of “positive feelings". For Topic 93, we can see that there was a clear focus on "people, age, and COVID issues" with the top words being "covid", "young", "risk", "fever", "immune", "age", "sick", "cough", "life", "cold", "elderly", and "older". The terms “covid”, “young”, and “risk” were the most highlighted words, with word-weights of 0.0299\%, 0.0222\%, and 0.0218\%, respectively, and this topic had negative polarity.

\begin{figure}[h!]
  \centering
  \begin{subfigure}[b]{0.6\linewidth}
    \includegraphics[width=\linewidth]{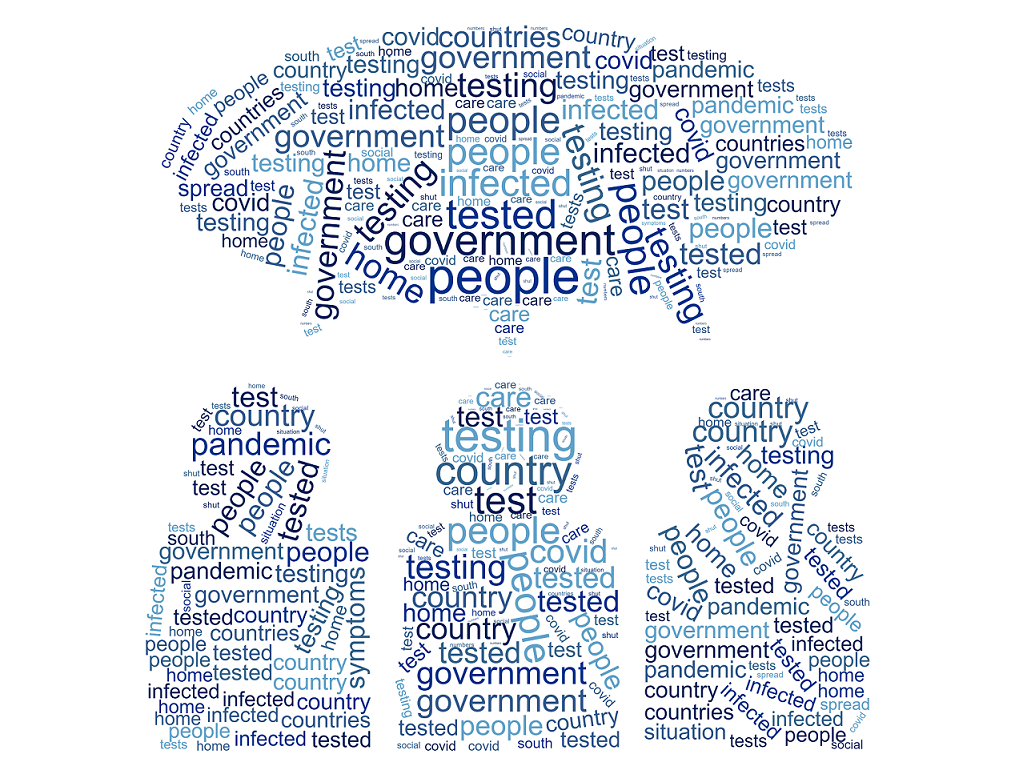}
    \caption{Topic Word 48}
  \end{subfigure}
  \begin{subfigure}[b]{0.6\linewidth}
    \includegraphics[width=\linewidth]{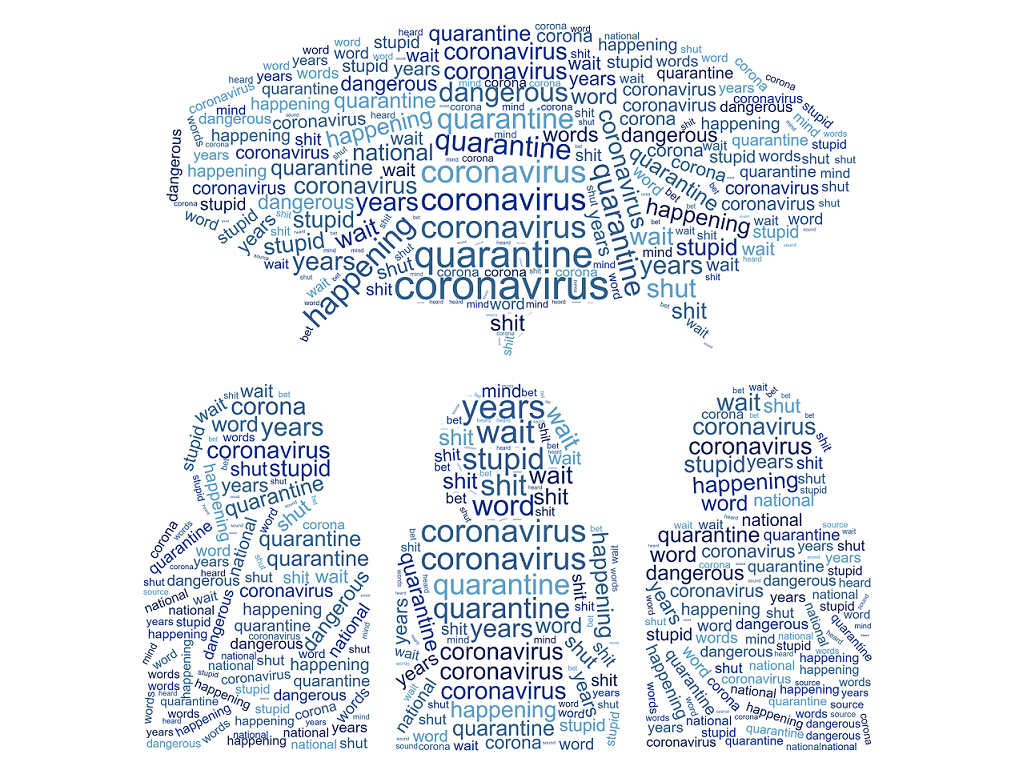}
    \caption{Topic Word 17}
  \end{subfigure}
  \caption{Word cloud visualisation based on the word-weight of the topics.}
  \label{fig:coffee}
\end{figure}

Topic 48 also addresses "COVID-19 testing issues" and contains words like "people", "testing", "government", "country", "tested", "test", "infected", "home", "covid", and "pandemic". Based on the results, the terms "people" and "testing" were the most highlighted words with word weights of 0.0447\% and 0.0337\%, respectively. Moreover, the opinion words based on sentiment analysis scored high in negative polarity for Topic 17. The top terms of this topic were “coronavirus", "quarantine", "stupid", "happening", "shit", "watch", and "dangerous", thus pertaining to the phenomenon "quarantine issues". The terms "coronavirus" and "quarantine" were the most highlighted words, with word-weights of 0.0353\% and 0.0346\%, respectively.

\subsection{Sentiment and Polarity results}
Sentiment analysis is a practical technique in NLP for opinion mining that can be used to classify text/comments based on word polarities [28] - [30]. This technique has many applications in various disciplines, such as opinion mining in online healthcare communities [31] - [33]. We obtained the sentiment of the COVID-19--related comments using the SentiStrength algorithm [34] - [36]. Therefore, with all COVID-19--related comments tagged with sentiment scores, we calculated the average sentiment of the entire dataset along with comments mentioning only 10 COVID-19 sub-reddits. The main objective of this analysis was to identify the overall sentiment of the COVID-19--related comments. We calculated the average sentiment of all comments as negative, positive, or neutral. Figure 9 shows the sentiment of all comments in the database along with the average sentiment of comments containing the terms COVID-19.

\begin{figure*}
\centering
  \includegraphics[height=7.12cm,width=10cm]{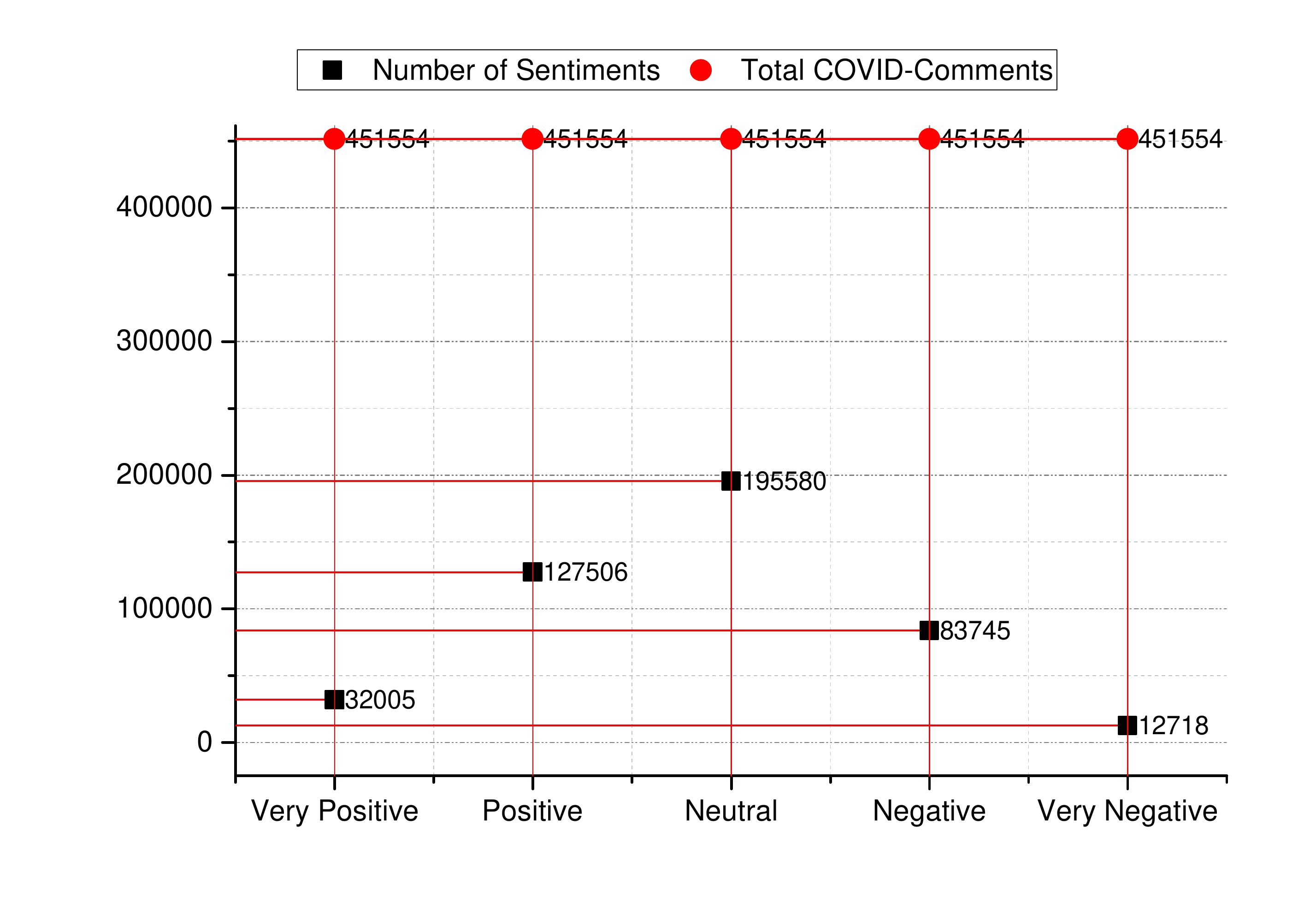}
\caption{ Distribution of COVID-19 comments with positive, negative or neutral sentiments of reddit Data}
\label{fig:1}       
\end{figure*}


\begin{landscape}
\begin{table}[]
\caption{Examples of COVID-19 comments from the reddit corpus}

\resizebox{19cm}{!} {
\begin{tabular}{|l|l|l|}
\hline
\multicolumn{1}{|c|}{Polarity}     & \multicolumn{1}{c|}{People's Comment}                                                                                                                          & \multicolumn{1}{c|}{Score of the words}                                                                                                                                                                                                                                                                                                                                                                             \\ \hline
\multirow{3}{*}{\textbf{Positive}} & I hope loved ones remain safe healthy.                                                                                                                         & \begin{tabular}[c]{@{}l@{}}I{[}0{]} hope{[}2{]} loved{[}3{]}{[}+1 MultiplePositiveWords{]} ones{[}0{]} remain{[}0{]} safe{[}1{]} healthy{[}0{]} \\ {[}{[}Sentence=-1,5=word max, 1-5{]}{]}{[}{[}{[}5,-1 max of sentences{]}{]}{]}\end{tabular}                                                                                                                                                                       \\ \cline{2-3} 
                                   & \begin{tabular}[c]{@{}l@{}}Ah yes manbaby magnificent immune system. \\ better luck next time covid-19\end{tabular}                                            & \begin{tabular}[c]{@{}l@{}}Ah{[}0{]} yes{[}0{]} manbaby{[}0{]} magnificent{[}3{]}\\   immune{[}0{]} system{[}0{]} {[}{[}Sentence=-1,4=word max, 1-5{]}{]} better{[}0{]} luck{[}2{]} next{[}0{]}\\   time{[}0{]} covid{[}0{]} 19{[}0{]} {[}{[}Sentence=-1,3=word max, 1-5{]}{]}{[}{[}{[}4,-1 max of\\   sentences{]}{]}{]}\end{tabular}                                                                               \\ \cline{2-3} 
                                   & \begin{tabular}[c]{@{}l@{}}I really hope whole eu follows italy. \\ they shut everything pandemic .\end{tabular}                                               & \begin{tabular}[c]{@{}l@{}}I{[}0{]} really{[}0{]} hope{[}2{]}{[}1LastWordBoosterStrength{]} whole{[}0{]} eu{[}0{]} follows{[}0{]} italy{[}0{]} \\ {[}{[}Sentence=-1,4=word max, 1-5{]}{]} they{[}0{]} shut{[}0{]} everything{[}0{]} pandemic{[}0{]}\\  {[}{[}Sentence=-1,1=word max, 1-5{]}{]}{[}{[}{[}4,-1 max of sentences{]}{]}{]}\end{tabular}                                                                   \\ \hline
\multirow{3}{*}{\textbf{Negative}} & Greed prejudice racism hate kill faster covid-19 .                                                                                                             & \begin{tabular}[c]{@{}l@{}}greed{[}-2{]} prejudice{[}-2{]}{[}-1\\   MultiplePositiveWords{]} racism{[}-1{]} hate{[}-3{]} kill{[}-1{]} faster{[}0{]} covid{[}0{]} 19{[}0{]}\\   {[}{[}Sentence=-4,1=word max, 1-5{]}{]}{[}{[}{[}1,-4 max of sentences{]}{]}{]}\end{tabular}                                                                                                                                           \\ \cline{2-3} 
                                   & \begin{tabular}[c]{@{}l@{}}" Deeply concerned by inaction over the virus "- because \\ you refused to identify this as a pandemic fucking fucks .\end{tabular} & \begin{tabular}[c]{@{}l@{}}deeply{[}0{]} concerned{[}-1{]} by{[}0{]} inaction{[}0{]}\\   over{[}0{]} the{[}0{]} virus{[}0{]} because{[}0{]} you{[}0{]} refused{[}-1{]} to{[}0{]} identify{[}0{]}\\   this{[}0{]} as{[}0{]} a{[}0{]} pandemic{[}0{]} fucking{[}0{]} fucks{[}-2{]}{[}-2\\   LastWordBoosterStrength{]} {[}{[}Sentence=-5,1=word max, 1-5{]}{]}{[}{[}{[}1,-5 max of\\   sentences{]}{]}{]}\end{tabular} \\ \cline{2-3} 
                                   & So much bullshit one thread alone. scary times .                                                                                                               & \begin{tabular}[c]{@{}l@{}}so{[}0{]} much{[}0{]} bullshit{[}-2{]} one{[}0{]} thread{[}0{]}\\   alone{[}0{]} {[}{[}Sentence=-3,1=word max, 1-5{]}{]} scary{[}-3{]} times{[}0{]}\\   {[}{[}Sentence=-4,1=word max, 1-5{]}{]}{[}{[}{[}1,-4 max of sentences{]}{]}{]}\end{tabular}                                                                                                                                       \\ \hline
\multirow{3}{*}{\textbf{Neutral}}  & I heard radio likely official guidance next 10 - 14 days                                                                                                       & \begin{tabular}[c]{@{}l@{}}i{[}0{]} heard{[}0{]} radio{[}0{]} likely{[}0{]} official{[}0{]}\\   guidance{[}0{]} next{[}0{]} 10{[}0{]} 14{[}0{]} days{[}0{]} {[}{[}Sentence=-1,1=word max,\\   1-5{]}{]}{[}{[}{[}1,-1 max of sentences{]}{]}{]}\end{tabular}                                                                                                                                                          \\ \cline{2-3} 
                                   & Everyone wear mask case unintentionally spreading everyone .                                                                                                   & \begin{tabular}[c]{@{}l@{}}everyone{[}0{]} wear{[}0{]} mask{[}0{]} case{[}0{]}\\   unintentionally{[}0{]} spreading{[}0{]} everyone{[}0{]} {[}{[}Sentence=-1,1=word max,\\   1-5{]}{]}{[}{[}{[}1,-1 max of sentences{]}{]}{]}\end{tabular}                                                                                                                                                                           \\ \cline{2-3} 
                                   & Would kind enough link one studies stories ?                                                                                                                   & \begin{tabular}[c]{@{}l@{}}would{[}0{]} kind{[}1{]}{[}-1 LastWordBoosterStrength{]}\\   enough{[}0{]} link{[}0{]} one{[}0{]} studies{[}0{]} stories{[}0{]} {[}{[}Sentence=-1,1=word max,\\   1-5{]}{]}{[}{[}{[}1,-1 max of sentences{]}{]}{]}\end{tabular}                                                                                                                                                           \\ \hline
\end{tabular}
}
\end{table}
\end{landscape}

For each of the polar comments in our labelled dataset, we assigned negative and positive scores utilizing SentiStrength, and employed the various scores directly as rules for building inference about the polarity/sentiment of the COVID-19 comments. Based on SentiStrength, we determined that a comment was positive if the positive sentiment score was greater than the negative sentiment score, and also considered a similar rule for determining a positive sentiment. For example, a score of +5 and -4 indicates positive polarity and a score of +4 and -6 indicates negative polarity. Moreover, If the sentiment scores were equal (such as -1 and +1, +4 and -4), we determined that the comment was neutral.

\subsection{Deep classification and Feature Analysis}

To prepare the dataset to automatically classify the sentiment of the COVID-19 comments for all of the data, we labelled each of the comments as very positive, positive, very negative, negative, and neutral based on the sentiment score obtained using the Sentistrength method. The training set had 338,666 COVID-19--related comments and the testing set had 112,888 comments. In this experiment, we evaluated the proposed LSTM-model and also supervised machine-learning methods using the Support Vector Machine (Senti-ML1), Naive Bayes (Senti-ML2), Logistic Regression (Senti-ML3), K Nearest Neighbors (Senti-ML4) techniques. Figure 4 shows the accuracy of the best model for classifying a COVID-19 comment as either a very positive, positive, very negative, negative, or neutral sentiment. Our approach based on the LSTM model, which classified all COVID-19 comments in the majority class achieved 81.15\% accuracy, which was higher than that of traditional machine-learning algorithms. We believe that the sentiment and semantic techniques can provide meaningful results with an overview of how users/people feel about the disaster.

\begin{figure*}
\centering
  \includegraphics[height=8.12cm,width=11cm]{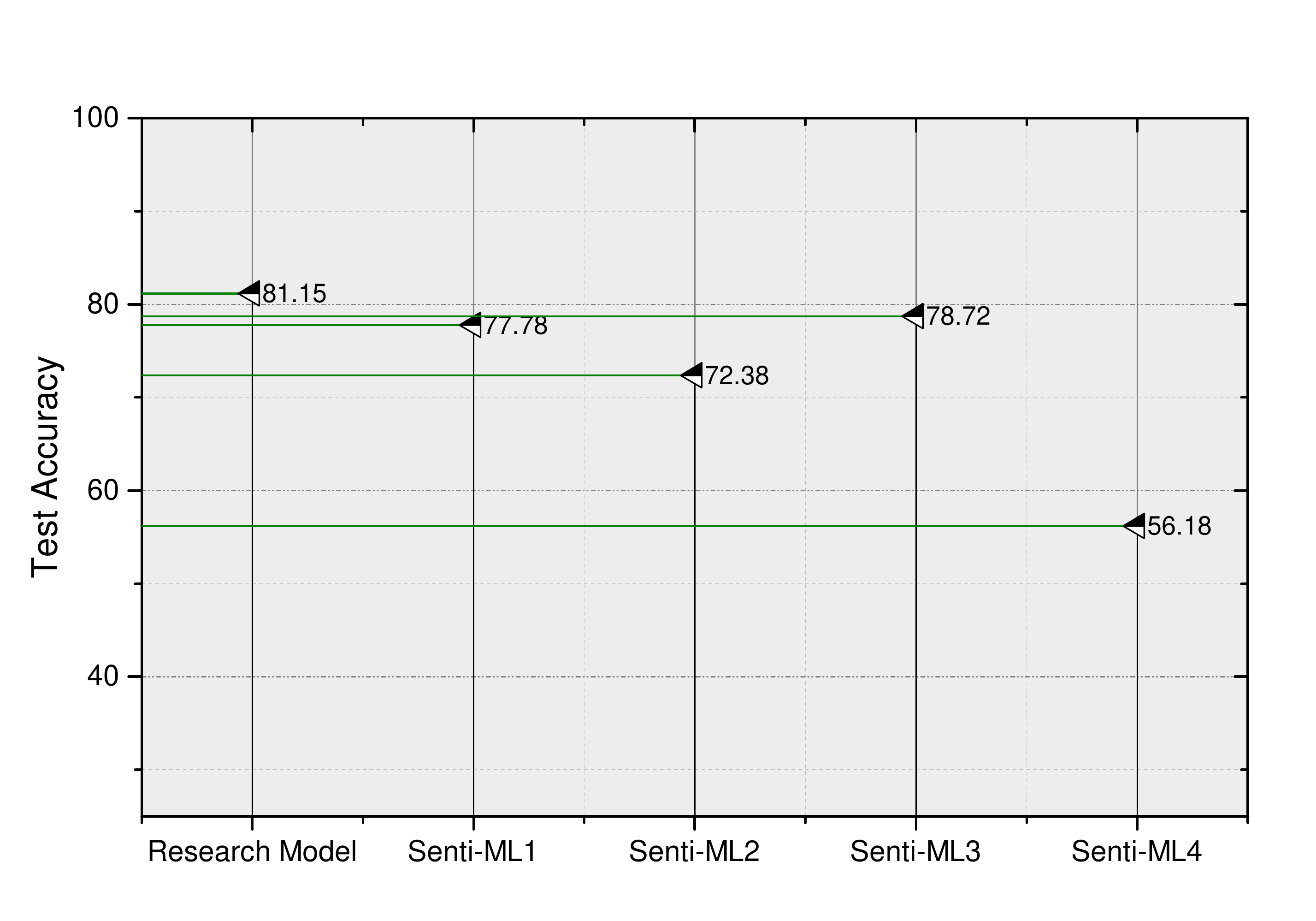}
\caption{ Accuracy performance of the methods for COVID-19 sentiment-classification using various features}
\label{fig:1}       
\end{figure*}

\section{Discussion and Practical Findings }

Analysing social media comments on platforms such as reddit could provide meaningful information for understanding people’s opinions, which might be difficult to achieve through traditional techniques, such as manual methods. The text content on reddit has been analysed in various studies [37] - [39]; to the best of our knowledge, this is the first study to analyse comments by considering semantic and sentiment aspects of COVID-related comments from reddit for online health communities.

Overall, we extended the analysis to check whether we could find a dependency of semantic aspects of user-comments for different issues on COVID-19--related topics. In this case, we considered an existing dataset that included 563,079 comments from 10 sub-reddits. We found and detected meaningful latent topics of terms about COVID-19 comments related to various issues. Thus, user comments proved to be a valuable source of information, as shown in Tables 1 and 2 and Figures 4-8. A variety of different visualisations was used to interpret the generated LDA results. As mentioned, LDA is a probabilistic model that, when applied to documents, hypothesises that each document from a collection has been generated as a mixture of unobserved (latent) topics, where a topic is defined as a categorical distribution over words. Regarding the top-ranked topics for the COVID-19 comments, it is possible to recognise many words probably related to needs and highlight-discussions of the people or users on reddit.

This research was limited to English-language text, which was considered a selection criterion. Therefore, the results do not reflect comments made in other languages. In addition, this study was limited to comments retrieved from January 20, 2020 and March 19, 2020. Therefore, the gap between the period in which the research was being completed and the time-frame of our study may have somewhat affected the timeliness of our results. Overall, the study suggests that the systematic framework by combining NLP and deep-learning methods based on topic modelling and an LSTM model enabled us to generate some valuable information from COVID-19--related comments. These kinds of statistical contributions can be useful for determining the positive and negative actions of an online community, and to collect user opinions to help researchers and clinicians better understand the behaviour of people in a critical situation. Regarding future work, we plan to evaluate other social media, such as Twitter, using hybrid fuzzy deep-learning techniques [40] - [41] that can be used in the future for sentiment level classification as a novel method of retrieving meaningful latent topics from public comments.

\section{CONCLUSION}

To our knowledge, this is the first study to analyse the association between COVID-19 comments’sentiment and semantic topics on reddit. The main goal of this paper, however, was to show a novel application for NLP based on an LSTM model to detect meaningful latent-topics and sentiment-comment-classification on COVID-19--related issues from healthcare forums, such as sub-reddits. We believe that the results of this paper will aid in understanding the concerns and needs of people with respect to COVID-19--related issues. Moreover, our findings may aid in improving practical strategies for public health services and interventions related to COVID-19.

\section*{Acknowledgements}
We acknowledge SciTechEdit International, LLC (Highlands Ranch, CO, USA) for providing pro bono professional English-language editing of this article. This work has been awarded by the National Natural Science Foundation of China (61941113, 81674099, 61502233), the Fundamental Research Fund for the Central Universities (30918015103, 30918012204), Nanjing Science and Technology Development Plan Project (201805036), and "13th Five-Year" equipment field fund (61403120501), China Academy of Engineering Consulting Research Project(2019-ZD-1-02-02).

\section*{Ethical Approval }
All procedures performed in studies involving human participants were in accordance with the ethical standards of the institutional and/or national research committee and with the 1964 Helsinki declaration and its later amendments or comparable ethical standards.\\\\\textbf{Declaration of Conflict of Interest :} All authors declare no conflict of interest directly related to the submitted work.\\

 \nocite{*}


\end{document}